# PHANTOM: Physics-Aware Adversarial Attacks against Federated Learning-Coordinated EV Charging Management System

Mohammad Zakaria Haider
Florida International University
Miami, Florida, USA
mhaid010@fiu.edu

Amit Kumar Podder
North Carolina State University
Raleigh, North Carolina, USA
apodder@ncsu.edu

Prabin Mali
Florida International University
Miami, Florida, USA
pmali004@fiu.edu

Aranya Chakrabortty
North Carolina State University
Raleigh, North Carolina, USA
achakra2@ncsu.edu

Sumit Paudyal
Florida International University
Miami, Florida, USA
spaudyal@fiu.edu

Mohammad Ashiqur Rahman
Florida International University
Miami, Florida, USA
marahman@fiu.edu

## Abstract

The rapid deployment of electric vehicle charging stations (EVCS) within distribution networks necessitates intelligent and adaptive control to maintain the grid's resilience and reliability. In this work, we propose PHANTOM, a **ph**ysics-aware **a**dversarial **n**etwork through **t**raining and **o**ptimization of **m**ulti-agent reinforcement learning model. PHANTOM integrates a physics-informed neural network (PINN) enabled by federated learning (FL) that functions as a digital twin of EVCS-integrated systems, ensuring physically consistent modeling of operational dynamics and constraints. Building on this digital twin, we construct a multi-agent RL environment that utilizes deep Q-networks (DQN) and soft actor-critic (SAC) methods to derive adversarial false data injection (FDI) strategies capable of bypassing conventional detection mechanisms. To examine the broader grid-level consequences, a transmission–distribution (T&D) dual simulation platform is developed, allowing us to capture cascading interactions between EVCS disturbances at the distribution level and the operations of the bulk transmission system. Results demonstrate how learned attack policies disrupt load balancing and induce voltage instabilities that propagate across T&D boundaries. These findings highlight the critical need for physics-aware cybersecurity to ensure the resilience of large-scale vehicle-grid integration.

## 1 Introduction

### 1.1 Motivation

The electrification of transportation has witnessed unprecedented growth in recent years, with electric vehicle charging stations (EVCSs) emerging as a critical component of the electric vehicle (EV) ecosystem. Integrating these charging stations into the power grid introduces a new dimension of cyber-physical interdependence, underscoring the necessity for robust cybersecurity measures. Among the myriad threats facing EVCS, false data injection (FDI) attacks pose a particularly insidious risk, potentially compromising the integrity and functionality of these critical systems. Recently, power grids have become smarter and more efficient by incorporating the Internet of Things (IoT), making them vulnerable to cyberattacks. A recent report found that the California power grid has defended over a million cyber-attacks each month [1]. FDI attacks represent a sophisticated cyber threat wherein adversaries manipulate data within a system to deceive its decision-making processes. In the context of EVCS, these attacks can lead to incorrect charging parameters, affecting the state of charge (SoC) and current reference values, which consequently compromise the charging process efficiency and potentially cause damage to the EVs.

Trends in electrification suggest that one in three cars is expected to be electrified by 2040 [2]. The current EV charging infrastructure has to be improved and expanded in accordance with the global auto fleet's shift toward EVs. The motivations behind cyberattacks on an EVCS range from identity theft and electricity theft to ransomware and virus assaults that potentially compromise the entire EVCS network [3, 4]. The transition of the attack vector from the cyber layer to the physical infrastructure layer involves intricate metrics that should be analyzed in relation to the aftermath in real physical entities, such as power, current, voltage, and SoC. Soltan et al. in [5] demonstrated how high-wattage devices can disrupt the power grid. The work in [6, 7] includes a vulnerability analysis and risk assessment of an EVCS, providing details of potential attack scenarios, such as denial-of-service (DoS), man-in-the-middle (MiTM), and FDI. Ting et al. [8] demonstrated that the abundance of EVs can be leveraged to enhance the stability of the power grid. The adverse interaction between EVCSs and power grids has been presented in [9–11].

Cyber-physical system (CPS) security research for EVCS is hindered by limited access to operational models, ethical constraints on real-world experimentation, and the tightly coupled nature of modern power grids, where disturbances can cascade across transmission and distribution layers. Conventional simulation models often struggle to capture these cascading effects due to issues with data quality and the complexity of nonlinear interactions among grid components. Physics-informed neural networks (PINNs) address part of this gap by embedding governing equations into the learning process, enforcing conservation laws and operational constraints through physics-informed loss functions, and reducing reliance on extensive historical data. Extending this with federated learning across multiple distribution systems enables collaborative model training at geographically dispersed nodes while preserving data privacy and improving convergence. Integrating reinforcement learning (RL)-driven adaptive FDI attacks into this framework further exposes evolving adversarial behaviors and system weak points. Collectively, the combination of PINNs, federated optimization, and RL-based attack modeling yields a more realistic and comprehensive view of CPS security risks in EVCS.

## 1.2 Related Work

The ever-expanding electric vehicle infrastructure has prompted various approaches to manage the penetration of electric vehicles on the power grid. A charging management system (CMS) enables the monitoring of charging station activity, including charging, scheduling, and load balancing. Cyber-physical security challenges in extreme fast charging (XFC) stations for EVs, focusing on potential cyber threats that could destabilize charging networks and impact grid stability, which have been demonstrated in [12–16]. By injecting falsified data into demand forecasts or measurement values, attackers can disrupt operations within the EV charging ecosystem, affecting demand management and grid stability. The authors in [17] have modeled the EV charging network as a cyber-physical system and propose a charging station recommendation algorithm to spatially and temporally distribute charging loads. A coordinated switching attack was introduced there that leveraged the charging and discharging capabilities of EVs to destabilize the power grid. Simulations on power grid models, including the 39-bus New England system, showed that such attacks could induce oscillations capable of causing grid instability [18]. Basnet et al. in [19] explored cybersecurity issues in EVCS supported by 5G, focusing on vulnerabilities to FDI and DoS attacks. By analyzing threats like spoofing, tampering, and DoS, the study established a comprehensive cybersecurity framework for EVCS.

Recent studies in [20–22] have demonstrated the utility of PINNs in modeling nonlinear dynamic equations across various domains. Their flexibility, scalability, and capability to approximate solutions in computationally efficient ways make PINNs an ideal tool for analyzing the intricate interactions between EVCS networks and power grids. The authors in [23–28] have discussed the challenges that PINNs address, including data scarcity, interpretability, and physical consistency, providing a roadmap for future research that leverages PINNs to improve power grid performance and resilience. Prior EVCS cybersecurity research has explored threat enumeration, attack impact, and defense strategies, but gaps persist in adaptive attack synthesis, physics-informed control modeling, and quantifying cascading transmission–distribution effects. This section reviews existing approaches and situates our federated LSTM–PINN framework within this state of the art. This section critically examines existing approaches and positions our federated long short-term memory (LSTM)-PINN framework within the current state of the art. Girdhar et al. [12] applied STRIDE-based threat modeling combined with weighted attack-defense trees to identify potential vulnerabilities in XFC stations. They utilize Hidden Markov Models (HMMs) to characterize attack probabilities and propose mitigations, primarily at the taxonomy level. However, their approach lacks adaptive target selection and does not quantify cascading impacts on distribution operations or transmission-level stability, leaving dynamic, learning-based adversarial behaviors against modern EVCS infrastructure unmodeled.

The disinformation attack paradigm explored by Pourmirza et al. [13] demonstrated how coordinated misinformation campaigns could influence energy consumption patterns in EVCS networks. However, their work provides only qualitative impact narratives without establishing a rigorous framework for attack propagation modeling or quantitative assessment of grid-level consequences. The absence of transmission-distribution co-simulation limits the practical applicability of their findings to real-world grid operations where EVCS integration occurs primarily at the distribution level. Sayed et al. [16] developed a feedback control theory-based attack synthesis using linear matrix inequalities (LMIs) and validated their approach on a two-area test system. Although mathematically rigorous, their analysis relies on linearized dynamics and assumes complete knowledge of the attacker's system parameters, generator characteristics, and controller settings, which limits realism and scalability to larger, more complex grids. It also overlooks anomaly-detection evasion and employs an oversimplified EV representation, treating 200 MW as bulk EV charging directly at the transmission level, rather than reflecting the actual siting of EVCS in distribution networks.

Acharya et al. [29] provided a comprehensive survey of cybersecurity challenges in EV-connected smart grids, cataloging various attack vectors and system vulnerabilities. The survey approach, while informative, does not advance beyond threat identification to provide actionable insights into attack timing, target selection, or impact mitigation. Khan et al. [30] demonstrated how compromised EV botnets and fast-charging stations could orchestrate coordinated load injection attacks on power grids. Their analysis revealed potential for significant load increases during targeted time windows; however, the attack strategies remained static and did not incorporate learning-based adaptation or stealth constraints. The work assumes simplified botnet coordination without considering the complexities of distributed EVCS deployment across multiple distribution feeders or the interaction with a sophisticated anomaly detection model (ADM).

Existing approaches share several limitations that our work addresses. Most assume static attack patterns and complete system knowledge, neglecting adaptive, learning-enabled adversaries. Their reliance on linearized models overlooks nonlinear, multi-scale interactions in EVCS-integrated grids; they rarely consider federated, privacy-preserving coordination, and they lack transmission–distribution co-simulation to trace how local EVCS compromises propagate to system-wide instability. In contrast, our PHANTOM framework utilizes an LSTM-enhanced PINN architecture that captures temporal dependencies and the underlying physics of power systems without linearization. A federated learning deployment enables privacy-preserving coordination across multiple distribution utilities while preserving consistent control logic. A multi-agent RL scheme couples a DQN for strategic target selection with a SAC agent for continuous, multi-signal FDI generation under stealth and physics-aware constraints. Finally, hierarchical transmission–distribution co-simulation explicitly links EVCS-level attacks to transmission-level stability risks, yielding a more realistic and comprehensive cyber-physical security assessment for EVCS-integrated power systems.

## 1.3 Contribution

This work presents PHANTOM, a physics-aware adversarial network that is trained and optimized through a multi-agent reinforcement learning model to systematically compromise federated LSTM-enhanced PINNs governing EVCS management in distribution networks. The contributions advance both adversarial machine learning and power system cybersecurity through the following technical innovations:

- We develop a novel LSTM-augmented PINN architecture that simultaneously captures temporal dependencies in EV charging patterns and enforces fundamental physical constraints of the power system. The PINN framework integrates Kirchhoff's laws, nodal power balance equations, linearized EVCS dynamics, and voltage stability requirements as differentiable constraints within the neural network training process, ensuring physical consistency while enabling rapid inference. The LSTM components provide predictive capabilities for non-stationary charging demands and grid conditions. This hybrid approach delivers significant computational advantages over conventional iterative power flow methods while maintaining rigorous adherence to power system physics, making it suitable for real-time charging management applications.
- We implement a horizontal federated learning approach in which a single, globally optimized LSTM-enhanced PINN charging optimizer is deployed identically across multiple autonomous charging management systems. No raw operational data is shared between systems; each node performs inference locally using its own system-specific features, preserving strict data locality. The federated setup maintains privacy by isolating all computations at the node level while achieving coordinated behavior through shared model parameters. This mirrors real-world multi-utility operations where consistency of control logic is needed without exposing proprietary data.
- We formalize the adversarial problem as a constrained decision-making process with hybrid actions and solve it using a coordinated DQN and SAC framework. The DQN network handles discrete structural choices, selecting which EVCS buses to target and the attack duration, while the SAC generates continuous measurement perturbations subject to predefined operational and stealth constraints. The objective balances two goals: maximizing physical and operational impact while minimizing detectability, with physics-aware constraints ensuring that all actions remain feasible within the system's operating limits.
- We develop a hierarchical T&D co-simulation environment comprising a 14-bus IEEE transmission system operating at high-voltage levels (approximately 138–500 kV) coupled with six independent 34-bus IEEE distribution feeders operating at medium-voltage levels (approximately 4–35 kV). The testbed employs a dual-engine architecture, comprising a Newton–Raphson power flow for transmission-level analysis and an OpenDSS solver for unbalanced distribution-level behavior and DER integration. This enables rigorous assessment of how EVCS phantom attacks propagate from distribution networks to impact transmission-level stability under realistic operational constraints.

The integrated framework demonstrates that coordinated, physics-aware adversarial learning can systematically exploit federated PINN-based charging optimization systems, inducing cascading instabilities across transmission-distribution interfaces while maintaining stealth against conventional anomaly detection mechanisms. These findings establish critical vulnerability baselines for next-generation cyber-resilient power system architectures. The rest of the paper is organized as follows: we provide a technical overview of the system dynamics in Section 2. Section 3 discusses the technical details of problem formalization. Section 4 explains the testbed setup, test cases, and evaluation. We have concluded the paper in Section 5.

# 2 TECHNICAL OVERVIEW

## 2.1 EVCS Dynamics

Podder et al. in [31] discussed the dynamics of EVCS modeling. In general, AC-to-DC power conversion for EV charging involves multiple stages to ensure safe and efficient operation. Grid synchronization, managed by a phase-locked loop (PLL), aligns the control system with the grid voltage by continuously adjusting for discrepancies in grid angle and frequency. The rectified AC power is stored in a DC link, where a control loop stabilizes the voltage by minimizing fluctuations caused by grid disturbances or load variations. Reference currents for the direct and quadrature AC components are computed for current regulation managed by an inner loop that compensates for component resistance and inductance. An LCL filter smooths AC currents and removes harmonics before rectification. The DC-link balances the power input from the grid with the EV's consumption, while a DC-DC converter modulates the stabilized voltage to match the EV battery's requirements. This ensures precise, reliable, and efficient charging, safeguarding battery health and maintaining overall system stability.

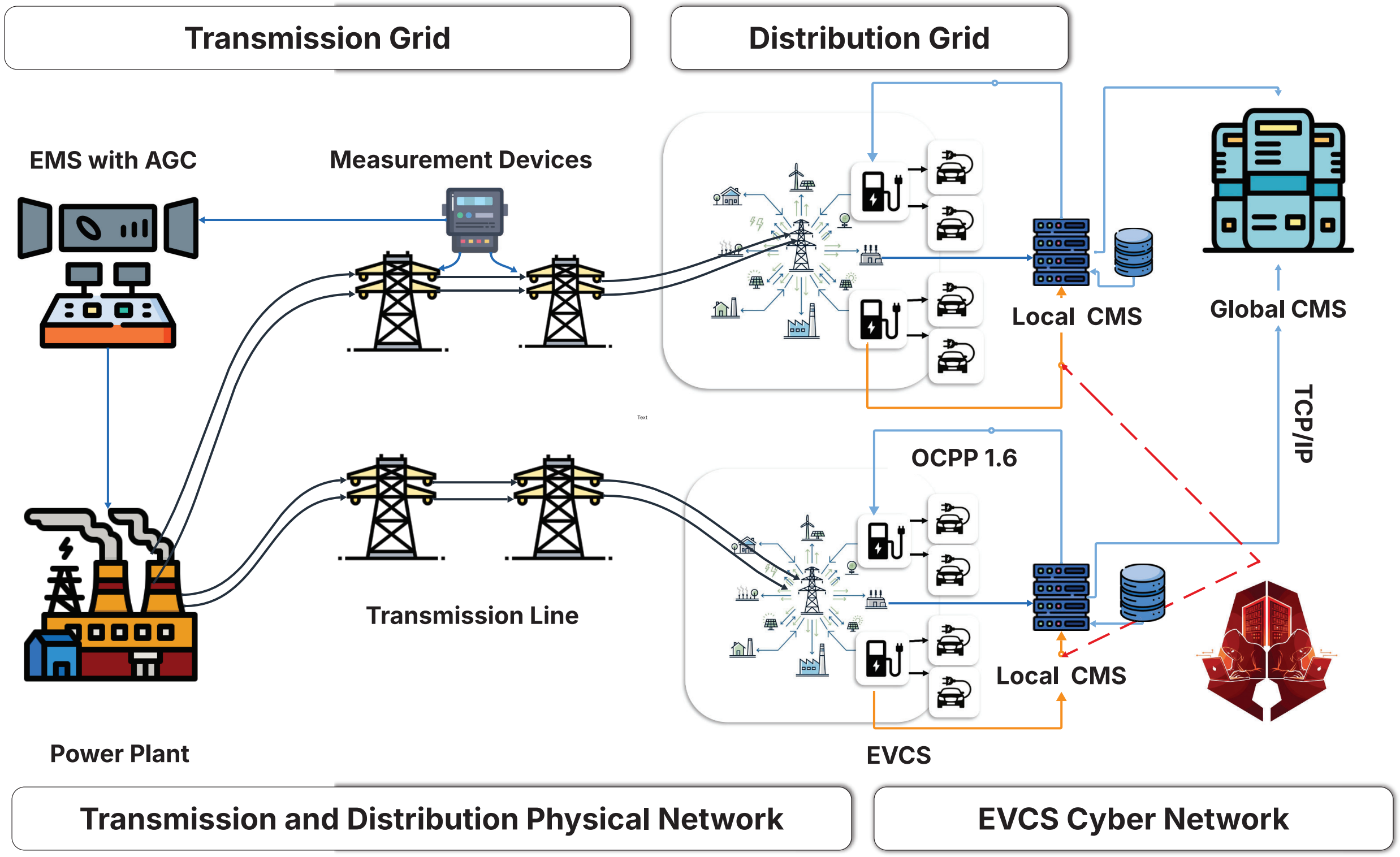

**Figure 1: Grid-connected EVCS network.**

## 2.2 LSTM-PINN Architecture and Training

Recent advances in scientific machine learning address the modeling bottlenecks of non-linear, multi-scale grid-cyber dynamics by embedding governing equations into neural architectures. Within this paradigm, PINNs eliminate explicit meshing and directly regularize learning with physical invariants, enabling accurate and efficient approximation of both forward and inverse dynamics in highly non-linear environments. In our setting, an LSTM-augmented PINN is used not only as a high-fidelity surrogate of EVCS–distribution dynamics, but as a local CMS that maps sensor streams and temporal context to optimal charging setpoints ($v_{ref}$, $i_{ref}$, $p_{ref}$) subject to operational limits and embedded physics. This yields real-time decision quality comparable to iterative optimization while preserving physical feasibility under non-stationary load and charging behavior. The proposed PINN serves as the core for charging optimization in EVCS operation. The PINN is trained with a composite objective that balances data fidelity and physics compliance as mentioned below:

$$\mathcal{L} = \lambda_d \, \mathcal{L}_{\text{data}} + \lambda_p \, \mathcal{L}_{\text{phys}} \tag{1}$$

with empirical weights $\lambda_d = 1.0$ and $\lambda_p = 10$. The data term aligns model outputs with trajectories generated from IEEE 34-bus distribution simulations and realistic EVCS parameters:

$$\mathcal{L}_{\text{data}} = \frac{1}{N} \sum_{i=1}^{N} \left\| \hat{\mathbf{y}}_i - \mathbf{y}_i \right\|_2^2. \tag{2}$$

The physics-informed term penalizes violations of electrical and operational laws:

$$\mathcal{L}_{\text{phys}} = \mathcal{L}_{P=VI} + \mathcal{L}_{\text{SOC}} + \mathcal{L}_{\text{volt}} + \mathcal{L}_{\text{curr}} + \mathcal{L}_{\text{thermal}}, \tag{3}$$

where $\mathcal{L}_{P=VI}$ is active-power consistency that ensures $P - V{\cdot}I \rightarrow 0$, $\mathcal{L}_{\text{SOC}}$ calculates the deviation from feasible SOC-dependent charging profiles, $\mathcal{L}_{\text{volt}}$ ensures $V_{\text{min}} \leq V \leq V_{\text{max}}$, and $\mathcal{L}_{\text{curr}}$ keeps $I_{\text{min}} \leq I \leq I_{\text{max}}$, and $\mathcal{L}_{\text{thermal}}$ represents the thermal-efficiency and loss-avoidance penalties.

Boundary conditions from realistic charging sessions are imposed as auxiliary constraints to improve generalization. We pretrain on linearized EVCS–grid dynamics, validate against nonlinear time-domain trajectories from differential-equation simulations, and train with Adam using an adaptive learning rate, gradient clipping, dropout, and early stopping. The resulting model generates real-time reference signals that adhere to device limits and embedded physics, enabling mesh-free and scalable EVCS optimization under dynamic grid conditions. We employ an LSTM-augmented PINN as a digital twin of the EVCS network, enabling faster and more scalable training and deployment than conventional Simulink-based simulations and optimization solvers, while preserving comparable dynamics; Ellinas et al. [32] report speed-ups of two to three orders of magnitude for PINN-based ODE solvers over standard methods. In our setting, the PINN is primarily trained by minimizing the residuals of the governing differential–algebraic equations and feasibility penalties that enforce power balance, voltage, and charger-level constraints, rather than on a large labeled dataset.

A small synthetic dataset, derived from realistic EVCS parameters and operating points, is used only to regularize training and encode typical operating ranges. The model is deployed in a federated configuration, where weights are shared across feeders, while raw measurements remain local to each CMS, thereby enhancing privacy and consistency of control. This physics-first scheme directly addresses data scarcity concerns, and since detailed dataset curation is not central to our attack analytics contribution, we keep it outside the scope of this work. Industry deployments already demonstrate the feasibility of neural-network–based EV charging control and planning: Tesla's Supercharger network [33] optimizes station loading and route guidance using real-time data, and GM employs machine-learning models for charger site selection [34].

## 2.3 Charging Management System

The CMS operates in discrete time with an update cycle of $\Delta t$ = 100 ms, executing a four-stage loop: local data collection, federated PINN optimization, reference application, and integration of power electronic dynamics. We use a 100 ms update step to match realistic EVCS control and sensing rates rather than for numerical convenience. Inner voltage–current loops in commercial chargers operate at frequencies of 10–100 Hz, while outer energy-management and coordination logic runs more slowly, consistent with a 100-ms supervisor step. EV battery and converter dynamics have time constants of tens of milliseconds, so sub-100-ms perturbations are largely filtered out at the supervisory layer. To handle potential adversarial fast transients, the anomaly detector applies rate-of-change limits, discarding updates whose magnitude changes by more than 10–50% within a 100 ms interval. The workflow ensures physics-consistent and privacy-preserving setpoint computation, as well as real-time actuation, across distributed EVCS stations. At each cycle, every EVCS station acquires local and feeder-level telemetry to construct a feature vector $\mathbf{x}_s(t)$:

$$\mathbf{x}_s(t) = \left[\mathrm{SOC}_s(t),\ V_{\mathrm{bus}}(t),\ f_{\mathrm{sys}}(t),\ d_{\mathrm{day}}(t),\ w_V(t),\ u_s(t),\ \boldsymbol{\phi}_s(t-1)\right]^\top, \quad (4)$$

where $\mathrm{SOC}_s$ is the state of charge, $V_{\mathrm{bus}}$ the per-unit bus voltage, $f_{\mathrm{sys}}$ the system frequency, $d_{\mathrm{day}}$ a normalized demand factor, $w_V$ a voltage-support priority weight, $u_s$ an urgency factor, and $\boldsymbol{\phi}_s$ the LSTM hidden state encoding temporal context. This stage corresponds to local CMS data acquisition per station. The shared LSTM-enhanced PINN, pre-trained and distributively deployed, maps $\mathbf{x}_s(t)$ to station-specific setpoints while embedding power-flow physics and operational constraints:

$$\mathbf{r}_s(t) = \left[v_{\mathrm{ref},s}(t),\ i_{\mathrm{ref},s}(t),\ p_{\mathrm{ref},s}(t)\right]^\top = f_{\mathrm{PINN}}^{\mathrm{LSTM}}\big(\mathbf{x}_s(t);\ \boldsymbol{\theta}^*\big), \quad (5)$$

with bounds enforced by device ratings and feeder limits:

$$v_{\min} \le v_{\mathrm{ref},s}(t) \le v_{\max}, \quad i_{\min} \le i_{\mathrm{ref},s}(t) \le i_{\max}, \quad (6)$$

$$p_{\min} \le p_{\mathrm{ref},s}(t) \le p_{\max}, \quad p_{\mathrm{ref},s}(t) = v_{\mathrm{ref},s}(t)\, i_{\mathrm{ref},s}(t). \quad (7)$$

Here, $\boldsymbol{\theta}^*$ denotes the globally optimized parameter set distributed identically to all CMS nodes (federated model sharing), while inference and data remain strictly local. The CMS dispatches the optimized references to the station's power-electronic controllers:

$$\big(v_{\mathrm{ref},s}(t),\ i_{\mathrm{ref},s}(t),\ p_{\mathrm{ref},s}(t)\big) \ \rightarrow\ C_s, \quad (8)$$

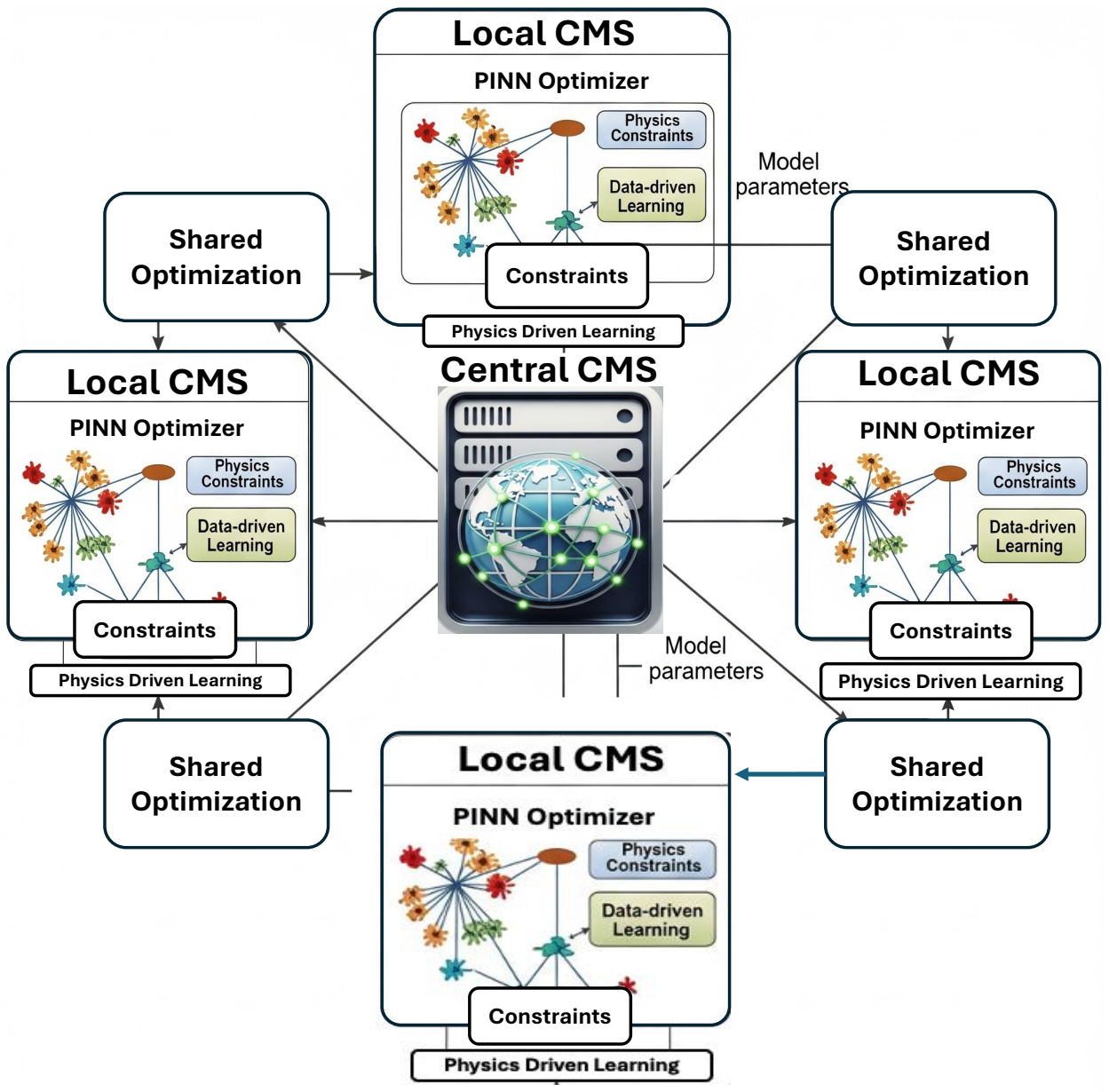

**Figure 2: Deep LSTM-based Federated PINN-based CMS.**

where $C_s$ denotes the local controller stack (outer power or voltage loop and inner current or voltage loop). The dq-frame control for the grid-side converter can be written as:

$$i_{d,\mathrm{ref}}(t) = \frac{p_{\mathrm{ref},s}(t)}{1.5\, v_d(t)}, \qquad i_{q,\mathrm{ref}}(t) \approx 0, \quad (9)$$

subject to current, thermal, and voltage constraints. Upon reference application, the station dynamics evolve according to the cascaded converter and charging subsystems. A simplified SOC update can be written as:

$$\mathrm{SOC}_s(t+\Delta t) = \mathrm{SOC}_s(t) + \frac{\eta_c\, i_{\mathrm{batt},s}(t)\, \Delta t}{C_{\mathrm{batt},s}}, \quad (10)$$

with charge efficiency $\eta_c$, battery current $i_{\mathrm{batt},s}$, and capacity $C_{\mathrm{batt},s}$. The feeder/node power balance and voltage constraints are implicitly enforced by the PINN's physics residuals during training and by runtime projection onto feasible sets. This loop yields real-time, physics-consistent setpoint generation that is privacy-preserving (with local data locality), scalable (utilizing shared model parameters), and robust to non-stationary operating conditions due to the LSTM temporal embedding.

## 2.4 Federated Learning-Enabled Central Charging Optimization

In large power systems, coordination across multiple, administratively distinct distribution networks must occur without exposing raw operational data. We therefore deploy the local PINN-CMS under a federated learning (FL) paradigm, sharing a common LSTM-augmented PINN model and parameters while keeping data local. This FL setup (i) preserves privacy and regulatory compliance, (ii) enforces consistent physics-informed control logic across heterogeneous feeders, and (iii) enhances scalability and resilience, as each node performs autonomous inference under intermittent communications. Consequently, coherent EVCS charging policies can be executed across $N$ autonomous distribution grids while respecting utility boundaries and local conditions. Consider grid-wide historical data $\mathcal{D}_{\text{global}} = \bigcup_{i=1}^{N} \mathcal{D}_i$, where $\mathcal{D}_i$ is the set of time-series measurements from grid $i$. The centralized training phase aims to find PINN model parameters $\boldsymbol{\theta}^* = \arg\min_{\boldsymbol{\theta}} \mathcal{L}_{\text{PINN}}(\mathcal{D}_{\text{global}}, \mathcal{P})$, with composite loss:

$$\mathcal{L}_{\text{PINN}} = \mathcal{L}_{\text{data}}(\boldsymbol{\theta}) + \lambda_{\text{phys}} \cdot \mathcal{L}_{\text{phys}}(\boldsymbol{\theta}, \mathcal{P}) \tag{11}$$

where $\mathcal{L}_{\text{data}}$ is the data-driven component, $\mathcal{L}_{\text{phys}}$ penalizes violations of physical constraints $\mathcal{P}$ (e.g., Kirchhoff's laws, voltage stability), and $\lambda_{\text{phys}}$ balances physical regularization. The trained model $f_{\text{PINN}}^{\text{LSTM}}(\mathbf{x}_t; \boldsymbol{\theta}^*)$ is broadcast to each CMS via a shared distribution mechanism:

$$\forall i \in \{1, \ldots, N\}: \quad \text{CMS}_i \longleftarrow f_{\text{PINN}}^{\text{LSTM}}(\cdot; \boldsymbol{\theta}^*)$$

thus ensuring all nodes receive identical model weights $\boldsymbol{\theta}^*$. Crucially, federated locality is maintained: no CMS $i$ shares its raw data $\mathcal{D}_i$ with any other node or a central server. The shared model parameters $\boldsymbol{\theta}^*$ are fixed during deployment, with all inference conducted locally:

$$\forall t,\ i: \quad \mathbf{y}_t^{(i)} = f_{\text{PINN}}^{\text{LSTM}}(\mathbf{x}_t^{(i)};\ \boldsymbol{\theta}^*)$$

where $\mathbf{y}_t^{(i)}$ are the local optimal setpoints at time $t$ for grid $i$. This federated LSTM-PINN framework achieves an identical optimizer with pretrained weights $\boldsymbol{\theta}^*$. The inputs $\mathbf{x}_t^{(i)}$ encode grid-specific features, ensuring appropriate optimization. There is no need for inter-grid sharing of measurements or states; new CMS nodes simply receive the pretrained model, and retraining is unnecessary. A single PINN-based optimizer is deployed across all EVCS instances, enabling each station to query the same surrogate in real-time to generate charging references that are consistent with nonlinear grid dynamics. This design eliminates localized training, reduces computational overhead, and preserves data locality and privacy while maintaining globally consistent, physics-aware control. During simulation, the PINN acts as a fast surrogate optimizer, enabling efficient evaluation of aggregate charging impacts without repeatedly solving large-scale optimization problems. Fig. 2 illustrates the federated learning module for the distributed CMS.

# 3 Attack Analytics Model

Conventional FDI impact studies discretize continuous, nonlinear grid dynamics via implicit integration (e.g., Backward Euler) and subsequently linearize interconnected components to couple a cyberattack model with anomaly-detection thresholds. While numerically stable, this pipeline introduces fidelity loss (higher-order/coupling effects are neglected), assumes static and omniscient adversaries (full knowledge of the plant and controller; targeted access to PMUs/RTUs/IEDs), and incurs prohibitive computational requirements for large systems, limiting real-time assessment and dynamic interplay between attackers and defenders. In contrast, multi-agent reinforcement learning (MARL) captures nonlinear, high-dimensional dynamics without linearization. It scales to support adaptive, long-horizon strategies, revealing insights into stealthy FDI sequences and mitigation policies. This approach yields a more robust and operationally relevant cyber-physical risk assessment.

## 3.1 Attack Technique

Following the MARL-based attack analytics described in Section 3.5, we now formulate a concrete FDI attack strategy against the CMS of EVCS-integrated distribution networks. The hybrid MARL framework, which combines DQN and SAC agents, jointly determines *where* to attack (target nodes, timing, and duration) and *how* to attack (magnitude and temporal profile). It injects malicious SOC, voltage, and demand measurements into the PINN optimizer to induce erroneous charging setpoints. Through sequential training and joint fine-tuning, the agents learn four main patterns—demand increase in low-load periods, demand reduction at peaks, oscillatory demand, and gradual ramp attacks—guided by stealth-aware reward shaping that favors large power/voltage deviations with low detectability. The RL agents never directly manipulate voltages or currents; they only falsify sensor inputs, causing the optimizer to make destabilizing decisions from corrupted data. Let $\tilde{\text{SOC}}_b^t$, $\tilde{V}_b^t$, $\tilde{P}_{d,b}^t$, and $\tilde{f}_b^t$ denote the adversarially injected components for state-of-charge, grid voltage, demand, and frequency at EVCS node $b$ and time step $t$, respectively. The resulting corrupted measurements, perceived by the CMS through the federated PINN optimizer, are given by:

$$\bar{\text{SOC}}_b^t = \text{SOC}_b^t + \tilde{\text{SOC}}_b^t, \tag{12}$$

$$\bar{V}_b^t = V_b^t + \tilde{V}_b^t, \tag{13}$$

$$\bar{P}_{d,b}^t = P_{d,b}^t + \tilde{P}_{d,b}^t, \tag{14}$$

$$\bar{f}_b^t = f_b^t + \tilde{f}_b^t, \tag{15}$$

for all $b \in \mathcal{B}$ and $t \in \mathcal{T}$, where $\mathcal{B}$ is the set of compromised EVCS buses and $\mathcal{T}$ is the attack time window.

(1) *State-of-Charge (SOC) Manipulation:*

$$\tilde{\text{SOC}}_b^t = \alpha_{\text{SOC}} \cdot \text{SOC}_b^t, \quad \alpha_{\text{SOC}} \leq 0.10$$

Artificially increasing SOC by up to 10% per unit, this attack deceives the optimizer into believing that batteries are more charged than they actually are, prompting premature termination or inappropriate curtailment of charging cycles.

(2) *Grid Voltage Falsification:*

$$\tilde{V}_b^t = -\alpha_V \cdot V_b^t, \quad \alpha_V = 0.02$$

Injecting voltage readings 2% lower than actual deceives the optimizer into overestimating voltage sags, possibly triggering unnecessary voltage regulation mechanisms.

(3) *Load/Demand Factor Falsification:*

$$\tilde{P}_{d,b}^t = -\alpha_P \cdot P_{d,b}^t, \quad 0.10 \leq \alpha_P \leq 0.15$$

By reducing the observed demand by 10–15% per unit, the system underestimates actual consumption, causing insufficient scheduling of charging resources and inaccurate power allocation.

(4) *Grid Frequency Manipulation:*

$$\tilde{f}_b^t = \alpha_f \cdot f_b^t, \quad \alpha_f = 0.001$$

Infusing frequency deviations of 0.1% per unit fabricates transient instability that can trick the optimizer into triggering frequency support or emergency control protocols.

The PINN-based federated optimizer thus receives, at each node and time:

$$\big(\bar{\text{SOC}}_b^t,\ \bar{V}_b^t,\ \bar{P}_{d,b}^t,\ \bar{f}_b^t\big), \tag{16}$$

and based on these corrupted signals, calculates suboptimal setpoints:

$$v_{\text{ref},b}^t = \mathcal{F}_v(\bar{\mathbf{x}}_b^t), \tag{17}$$

$$i_{\text{ref},b}^t = \mathcal{F}_i(\bar{\mathbf{x}}_b^t), \tag{18}$$

$$p_{\text{ref},b}^t = \mathcal{F}_p(\bar{\mathbf{x}}_b^t), \tag{19}$$

where $\bar{\mathbf{x}}_b^t = \big[\,\bar{\text{SOC}}_b^t,\ \bar{V}_b^t,\ \bar{P}_{d,b}^t,\ \bar{f}_b^t\,\big]^\top$ and $\mathcal{F}_v, \mathcal{F}_i, \mathcal{F}_p$ are the PINN-learned mappings for voltage, current, and power setpoint references, respectively. Multi-vector manipulations distort the CMS's state interpretation, causing unsafe charging schedules, voltage violations, and erroneous grid support. This accelerates asset degradation, compromising both local EVCS operations and global grid reliability.

## 3.2 Attack Constraints

The PHANTOM attack strategy is subject to two primary categories of constraints: system accessibility limitations and evasion of anomaly detection mechanisms.

**Accessibility:** Not every EVCS bus in the network is susceptible to compromise, due to inherent security measures or heterogeneous connectivity. Accessibility is modeled by a binary flag $\mathbb{A}_b$ assigned to each target bus $b$:

$$\forall b \in \mathcal{B}, \quad \mathbb{A}_b = 0 \implies \Delta\bar{\text{SOC}}_b^t = \Delta\bar{V}_b^t = \Delta\bar{P}_{d,b}^t = \Delta\bar{f}_b^t = 0, \tag{20}$$

where all injected deviations are zero for inaccessible buses.

**Stealth Constraints:** To evade ADM, agents constrain injected fluctuations within tolerable absolute and temporal bounds. For each bus $b$ and time $t$:

$$-\tau_{\text{SOC}} \le \bar{\text{SOC}}_b^{t+1} - \bar{\text{SOC}}_b^t \le +\tau_{\text{SOC}}, \tag{21}$$

$$-\tau_V \le \bar{V}_b^{t+1} - \bar{V}_b^t \le +\tau_V, \tag{22}$$

$$-\tau_P \le \bar{P}_{d,b}^{t+1} - \bar{P}_{d,b}^t \le +\tau_P, \tag{23}$$

$$-\tau_f \le \bar{f}_b^{t+1} - \bar{f}_b^t \le +\tau_f, \tag{24}$$

where $\tau_{\text{SOC}}$, $\tau_V$, $\tau_P$, $\tau_f$ are sensor-specific thresholds for inter-sample variability permitted by ADM.

**Operational Boundaries:** In addition, all corrupted measurements must be confined within system-defined operational boundaries:

$$\text{SOC}_b^{\min} \le \bar{\text{SOC}}_b^t \le \text{SOC}_b^{\max}, \tag{25}$$

$$V_b^{\min} \le \bar{V}_b^t \le V_b^{\max}, \tag{26}$$

$$P_{d,b}^{\min} \le \bar{P}_{d,b}^t \le P_{d,b}^{\max}, \tag{27}$$

$$f_b^{\min} \le \bar{f}_b^t \le f_b^{\max} \tag{28}$$

These constraints ensure physical plausibility and stealth, enabling the agent to maximize impact while minimizing detection in line with its reward structure. In our system, we have set maximum power injection limits of 45kW-55kW, voltage reference bounds of 350V-540V, frequency variation limits of up to 0.5Hz, current reference limits of 50A-150A, and rate-of-change thresholds for SOC of 10% per timestep for all charging parameters.

## 3.3 Multi-Stage ADM Architecture

To improve robustness, we employ a hierarchical three-layer defense-in-depth strategy tailored to EVCS networks under federated learning. The first layer employs a rule-based detection mechanism that performs real-time validation of incoming control signals against a comprehensive set of operational constraints specified in (25) through (28). This layer provides deterministic detection of blatant constraint violations with zero false positives. The second layer deploys an isolation forest-based statistical anomaly detector that analyzes multidimensional feature vectors comprising grid voltage, frequency, state of charge (SOC), demand factors, voltage priority, and urgency metrics. By constructing isolation trees that partition the feature space, this layer identifies attacks that manifest as statistical outliers from normal operational distributions, with a configurable anomaly threshold that balances detection sensitivity against false alarm rates. The third and most sophisticated layer utilizes an LSTM-based neural network with 128 hidden units, processing temporal sequences of 10 consecutive timesteps to capture complex attack signatures and temporal dependencies. Trained on benign data, the LSTM detects subtle patterns, such as drift, oscillations, and coordinated intrusions, that bypass earlier checks. All detection layers operate in parallel, with outputs aggregated via a logical OR to ensure maximum coverage.

## 3.4 Attack Assumptions

In order to evaluate practical risk scenarios and adversarial effectiveness, the following assumptions are adopted for the FLARE attack modeling:

- ***Assumption I:*** Lacking a complete system model, the attacker leverages MARL to incrementally learn manipulation strategies from local CMS/PINN feedback.
- ***Assumption II:*** Adversarial access is confined to real-time sensor streams at compromised EVCS nodes, precluding access to global grid controls or dispatch instructions.
- ***Assumption III:*** Attacks are confined to EVCS sensors via network exploits (e.g., ARP spoofing), enabling corruption that persists until interrupted by resets or integrity checks.
- ***Assumption IV:*** MARL agents exploit inherent system variability for adaptive attack timing, though security-driven resets restrict the attack window.
- ***Assumption V:*** Unaware of anomaly detection thresholds, the attacker learns stealth strategies through empirical trial and error.

These assumptions define a realistic adversary constrained to sensor-level manipulation and partial observability, operating within the practical limits of CMS-controlled EVCS networks. By explicitly linking optimization goals, constraints, and assumptions to the MARL formulation in Section 3.5, we establish a rigorous basis for analyzing physical and cyber vulnerabilities under coordinated FDI attacks.

### 3.5 Multi-Agent Reinforcement Learning Attack Analytics

To systematically characterize adversarial behaviors that can compromise federated PINN-based EVCS networks, we propose a coordinated multi-agent reinforcement learning (MARL) framework for synthesizing intelligent attacks. The framework explicitly models the interaction between adversarial strategies and evolving grid states, where dynamics are approximated by the LSTM-PINN surrogate to ensure physically realizable attack trajectories.

*3.5.1 Hybrid Agent Architecture.* The MARL framework utilizes two complementary agents that operate in a coordinated hybrid action space. A discrete DQN agent $\pi_D$ handles structural decisions by selecting target EVCS buses $b \in \mathcal{B}$ and attack duration classes $d \in \{1, \ldots, D_{\max}\}$, while a continuous SAC agent $\pi_C$ generates multi-signal perturbations across SOC, voltage, demand factor, and frequency measurements. The hybrid action at time $t$ is formulated as:

$$a_t^{(h)} = \mathcal{W}_h\big(a_t^{(D)}, a_t^{(C)}\big), \tag{29}$$

where $a_t^{(D)} \in \{0, \ldots, |\mathcal{B}|-1\} \times \{0, \ldots, D_{\max}-1\}$ encodes the target system and duration, and $a_t^{(C)} = [m_t, \theta_t] \in \mathbb{R}^2$ represents the magnitude and timing offset selected by the continuous agent.

*3.5.2 State Space and System Dynamics.* The system state incorporates physical grid observables, EVCS operational status, temporal components, and attack history:

$$\mathbf{s}_t = \big[L_t,\ V_t,\ f_t,\ I_t,\ \mathrm{SOC}_t,\ \sin(\omega t),\ \cos(\omega t),\ H_t\big], \tag{30}$$

where $L_t$ represents instantaneous load, $V_t, f_t, I_t$ denote voltages, frequencies, and currents, $\mathrm{SOC}_t$ captures sampled EVCS states-of-charge, and $H_t$ encodes recent adversarial magnitudes. The system evolution is governed by:

$$\mathbf{s}_{t+1} \sim P(\mathbf{s}_{t+1}|\mathbf{s}_t, a_t^{(h)}), \quad r_t = R_h(\mathbf{s}_t, a_t^{(h)}), \tag{31}$$

with physics-consistent state transitions provided by the PINN surrogate:

$$\tilde{\mathbf{s}}_{t+1} = f_{\mathrm{PINN}}^{\mathrm{LSTM}}\big(\tilde{\mathbf{s}}_t, a_t^{(h)}\big). \tag{32}$$

*3.5.3 DQN Agent Optimization.* The DQN agent optimizes target selection and duration planning through Q-function learning. The target selection objective maximizes expected cumulative rewards for directing attacks toward vulnerable EVCS nodes:

$$Q^{\mathrm{DQN}}(\mathbf{s}_t, a_t^{\mathrm{DQN}}) = \mathbb{E}_{\mathbf{s}_{t+1}}\left[r_t^{\mathrm{DQN}} + \gamma \max_{a_{t+1}^{\mathrm{DQN}}} Q^{\mathrm{DQN}}(\mathbf{s}_{t+1}, a_{t+1}^{\mathrm{DQN}})\right], \tag{33}$$

where $r_t^{\mathrm{DQN}}$ rewards successful identification of vulnerable nodes and optimal timing. The DQN optimization objective is:

$$\max_{a_t^{\mathrm{DQN}}} Q^{\mathrm{DQN}}(\mathbf{s}_t, a_t^{\mathrm{DQN}}). \tag{34}$$

*3.5.4 SAC Agent Optimization.* The SAC agent maximizes voltage and power deviations through continuous false data injection while maintaining stealth constraints. The attacker's reward is defined by cumulative deviations across targeted EVCS nodes:

$$r_t^{\mathrm{attacker}} = \sum_{b \in \mathrm{targets}} \big|V_{\mathrm{out},b} - V_{\mathrm{nominal}}\big| + \big|P_{\mathrm{out},b} - P_{\mathrm{nominal}}\big|. \tag{35}$$

The SAC objective maximizes expected cumulative rewards with entropy regularization:

$$J^{\mathrm{attacker}} = \mathbb{E}_{\pi^{\mathrm{attacker}}}\left[\sum_{t=0}^{T} \gamma^t r_t^{\mathrm{attacker}}\right], \tag{36}$$

subject to the soft Bellman equation:

$$\begin{aligned} Q^{\mathrm{attacker}}(\mathbf{s}_t, a_t^{\mathrm{attacker}}) &= \mathbb{E}_{\mathbf{s}_{t+1}}\Big[r_t^{\mathrm{attacker}} + \gamma \mathbb{E}_{a_{t+1}^{\mathrm{attacker}} \sim \pi^{\mathrm{attacker}}} \\ &\big[Q^{\mathrm{attacker}}(\mathbf{s}_{t+1}, a_{t+1}^{\mathrm{attacker}}) - \alpha \log \pi^{\mathrm{attacker}}(a_{t+1}^{\mathrm{attacker}} \mid \mathbf{s}_{t+1})\big]\Big], \end{aligned} \tag{37}$$

where $\alpha$ controls the entropy bonus for exploration.

*3.5.5 Coordinated Reward Design.* The hybrid reward function balances attack effectiveness, stealth preservation, and temporal vulnerability exploitation:

$$R_h = \sum_{q \in \{P,V,I\}} \omega_q \mathbb{I}(\Delta_q \geq \gamma_q) + \lambda_s \cdot \mathrm{Stealth}(\mathbf{s}_t, a_t^{(h)}) - \lambda_d \cdot \mathrm{Detect}(a_t^{(h)}), \tag{38}$$

where $\Delta_q$ represents percentage deviations in power $(P)$, voltage $(V)$, or current $(I)$ relative to detection thresholds $\gamma_q$. The stealth component rewards low-detectability actions:

$$\mathrm{Stealth}(\mathbf{s}_t, a_t^{(h)}) = \max\{0, (5.0 - m_t) \cdot \kappa\}, \tag{39}$$

while the detection penalty discourages conspicuous behavior that exceeds BDD thresholds.

*3.5.6 Joint Optimization Framework.* The complete MARL optimization integrates both agents through coordinated learning:

$$\max_{a_t^{\mathrm{DQN}}} Q^{\mathrm{DQN}}(\mathbf{s}_t, a_t^{\mathrm{DQN}}), \tag{40}$$

$$\max_{\pi^{\mathrm{attacker}}} J^{\mathrm{attacker}} = \mathbb{E}_{\pi^{\mathrm{attacker}}}\left[\sum_{t=0}^{T} \gamma^t r_t^{\mathrm{attacker}}\right], \tag{41}$$

subject to physics-aware constraints that ensure attack feasibility:

$$\mathrm{SOC}_{\min} \leq \bar{\mathrm{SOC}}_t \leq \mathrm{SOC}_{\max}, \tag{42}$$

$$V_{\min} \leq \bar{V}_t \leq V_{\max}, \tag{43}$$

$$|\bar{V}_t - \bar{V}_{t-1}| \leq \tau_V, \quad |\bar{f}_t - \bar{f}_{t-1}| \leq \tau_f, \tag{44}$$

where the final constraints ensure inter-sample smoothness to evade ADM mechanisms. This coordinated MARL framework enables learning of sophisticated attack policies that (i) adaptively select vulnerable targets and optimal timing through DQN-based structural decisions, (ii) generate stealthy, physics-consistent false data injection through SAC-based continuous control, and (iii) maximize grid impact while preserving detectability constraints through physics-informed reward design and LSTM-PINN state evolution. The overall optimization objective of the PHANTOM adversary, as guided by the learned MARL policies, is to maximize cumulative deviations in corrupted sensor measurements observed by the CMS, thereby driving the federated PINN optimizer toward suboptimal, grid-destabilizing decisions. The attack variables comprise the falsified injected state of charge, voltage, demand and frequency signals, denoted by $\tilde{\mathrm{SOC}}_b^t$, $\tilde{V}_b^t$, $\tilde{P}_{d,b}^t$, and $\tilde{f}_b^t$ in the bus $b$ and time $t$. The

**Algorithm 1** Hybrid MARL Attack Analytics: Training and Execution Workflow

**Require:** PINN surrogate $f_{\text{PINN}}$, hybrid wrapper $\mathcal{W}_h$, initial states $s_0$

```
Initialize DQN policy π_D, SAC policy π_C
Initialize replay buffers B_D, B_C
while training not converged do
    Observe current state s_t
    Sample discrete action a_t^(D) ~ π_D(s_t)  ▷ target and duration
    Sample continuous action a_t^(C) ~ π_C(s_t, a_t^(D))  ▷ magnitude and timing
    Combine a_t^(h) = W_h(a_t^(D), a_t^(C))
    Compute perturbed dynamics x̃_{t+1} = f_PINN(x̃_t, a_t^(h))
    Update grid state s_{t+1} and compute reward r_t = R_h(s_t, a_t^(h))
    Store transition (s_t, a_t^(D), a_t^(C), r_t, s_{t+1})
    Update π_D from B_D via Q-learning
    Update π_C from B_C via SAC gradients
end while
Execution Phase:
Deploy π_D, π_C to generate intelligent attack sequences
Wrapper synthesizes per-step actions {a_t^(h)} for adversarial co-simulation
```

optimization is formulated as:

$$\max_{\tilde{\text{SOC}},\tilde{V},\tilde{P}_d,\tilde{f}} \sum_{b\in\mathcal{B}}\sum_{t\in\mathcal{T}}\Big(\left|\tilde{\text{SOC}}_b^t-\text{SOC}_s\right|+\left|\tilde{V}_b^t-V_s\right|+\left|\tilde{P}_{d,b}^t-P_{d,s}\right|+\left|\tilde{f}_b^t-f_s\right|\Big) \tag{45}$$

where $\text{SOC}_s$, $V_s$, $P_{d,s}$, and $f_s$ are the nominal setpoints for state-of-charge, voltage, demand, and frequency, respectively. The MARL policies select injected values $\tilde{\text{SOC}}$, $\tilde{V}$, $\tilde{P}_d$, $\tilde{f}$ to maximize (45), coordinating across multiple EVCS nodes and time steps. To preserve stealth, the magnitude and temporal variation of injected signals are constrained such that resultant measurements remain within system operational bounds and below anomaly detection thresholds, as detailed in Section 3.2. This coordinated adversarial strategy ensures persistent degradation of CMS situational awareness and federated PINN optimization fidelity, while minimizing the risk of attack exposure. The overall training procedure is summarized in Algorithm 1.

# 4 Evaluation

## 4.1 Multi-Scale Power System Co-Simulation Framework

*4.1.1 Hierarchical Grid Architecture.* The system employs a hierarchical co-simulation framework that models realistic power grid operations across multiple voltage levels, integrating transmission-level control, distribution network behavior, and individual EV charging dynamics, as shown in Fig. 1. At the highest level, the transmission layer is modeled by an IEEE 14-bus system operating at high voltage (≈138–500 kV), maintaining a frequency of 60 Hz via automatic generation control that updates generation setpoints every 1 second based on the area control error. This layer is coupled at buses 4, 5, 7, 9, 10, and 13 to six independent IEEE 34-bus distribution feeders (4–35 kV), each hosting 10 EVCS at selected nodes. This multi-scale configuration captures the interactions between bulk transmission and EVCS-rich distribution networks, providing a comprehensive co-simulation testbed for assessing the impacts of EVCS integration.

*4.1.2 Dual Simulation Engine Architecture.* The framework uses domain-specific simulation engines tailored to each grid segment. *pandapower* [35], a Python-based analysis library, models the transmission system using Newton–Raphson power flow and optimal power flow solvers. In parallel, OpenDSS [36] simulates the distribution networks with dedicated support for unbalanced three-phase operation, voltage regulation, and DER integration. This dual-engine setup improves fidelity and efficiency by exploiting the specialized strengths of each tool.

*4.1.3 PINN Integration.* The system employs an LSTM-enhanced PINN that fuses deep learning with fundamental power system physics. This model acts as an intelligent charging optimizer, learning from historical grid data while enforcing constraints such as Kirchhoff's laws, power balance, and voltage limits. The LSTM captures temporal dependencies in charging behavior and grid evolution, enabling predictive, constraint-aware optimization. The LSTM–PINN is implemented in PyTorch.

*4.1.4 Federated Learning-Based Distributed Intelligence.* Each distribution level CMS acts as an autonomous federated learning node, learning optimal EVCS control policies without sharing raw operational data. This privacy-preserving architecture enables collaborative optimization across utilities while maintaining data sovereignty and mirroring real-world coordination practices; the federated model is implemented in Python using PyTorch.

*4.1.5 Advanced Reinforcement Learning Attack Modeling.* The cybersecurity assessment leverages Stable-Baselines3 to implement

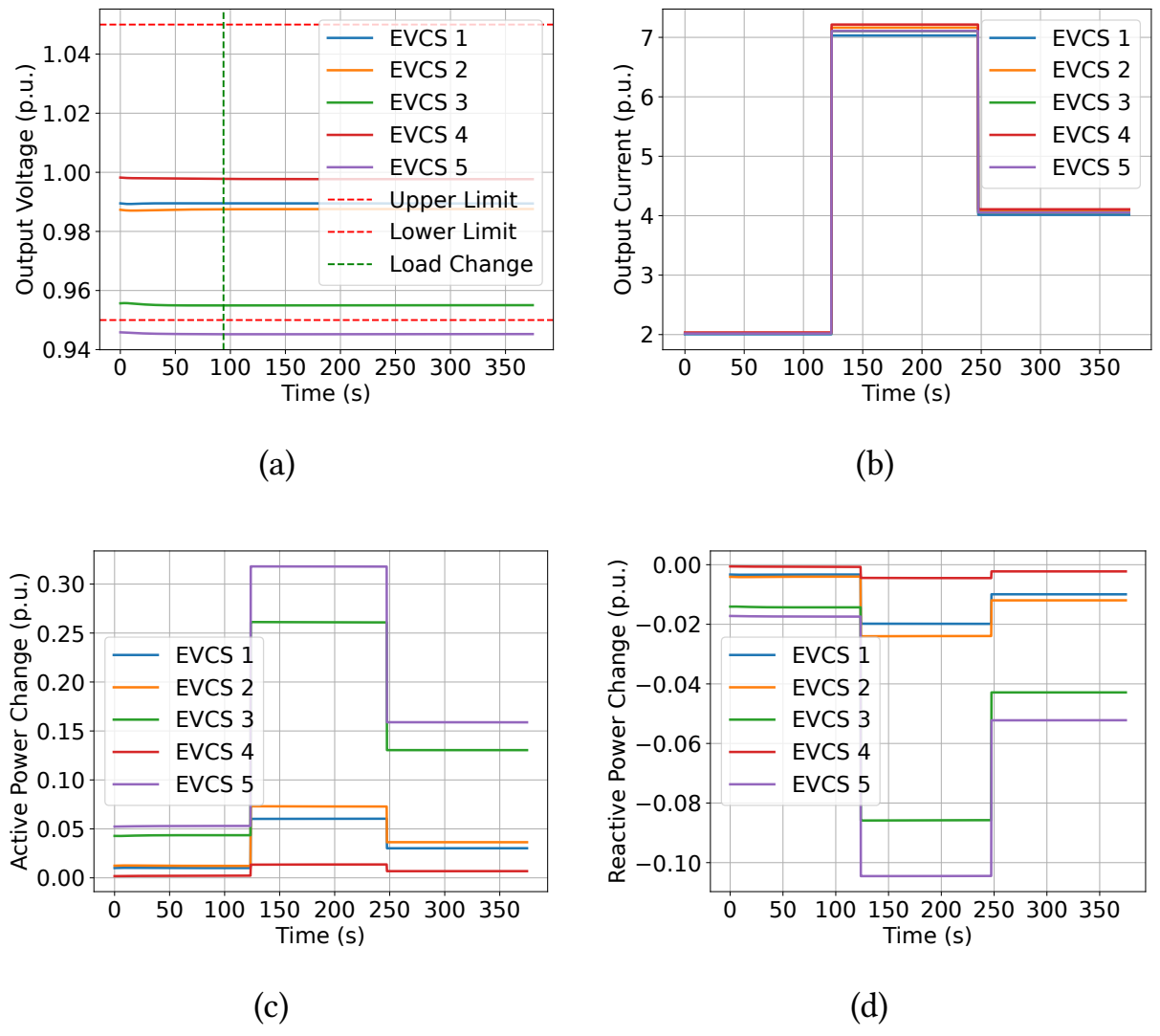

**Figure 3: Demonstrating outputs of EVCS (a) voltage, (b) current, (c) active power, and (d) reactive power in EVCS Bus.**

hybrid RL-based attack agents. DQN agents handle discrete strategic decisions (target selection, timing, duration), while SAC agents optimize continuous tactical parameters (e.g., attack magnitude, penetration level). Together, they generate realistic FDI attacks that corrupt sensor measurements supplied to the PINN optimizers, inducing AI-driven charging systems to make systematically suboptimal control decisions.

*4.1.6 Real-Time Co-Simulation Orchestration.* To ensure consistent state updates across transmission and distribution domains, the hierarchical co-simulation framework coordinates all components via time-synchronized steps. It supports dynamic load profiles, real-time attack injection, and multi-objective optimization over grid stability, charging efficiency, and cybersecurity resilience. This integration enables rigorous study of interactions between grid operations, AI-driven optimization, and cyber threats in a realistic multi-scale environment.

*4.1.7 Simulation Design.* The overall architecture has been developed in Python. The entire model runs on an Intel Xeon E5-240 platform equipped with two NVIDIA RTX 4090 GPUs and 248GB of RAM. The model comprises six individual IEEE 34-bus distribution grids, featuring ten EVCS stations with varying capacities ranging from 0.2 MW to 1 MW. The grid operates at a base power of 10 MVA, with nominal voltages in each layer. Each station has multiple EVCS ports designed with capacities of 50 kW and efficiencies of 98%, considering per-unit (p.u.) modeling. Before the full T&D co-simulation, we validated EVCS dynamics by applying up to 20% load variations at all EVCS-connected buses in the IEEE 34-bus system. As shown in Fig. 3, the CMS redistributes power and adjusts setpoints to keep voltages and power flows within operating limits, and the grid adapts smoothly to these perturbations. This confirms that the model realistically captures the interaction between EVCS demand variations and grid stability.

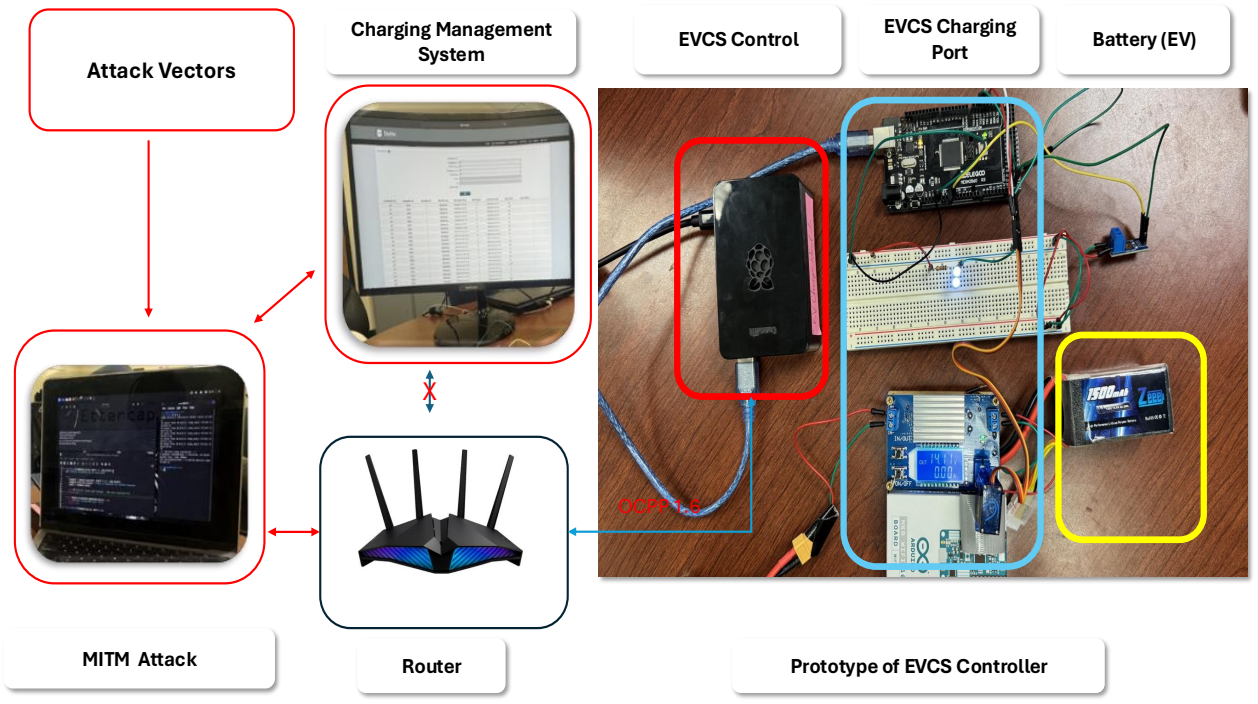

**Figure 4: Prototype of EVCS with OCPP Communication.**

*4.1.8 FDI Attack Injection via OCPP.* To validate the vulnerability of the OCPP module to FDI attacks, we built an EVCS prototype connected to a CMS via OCPP and executed a man-in-the-middle attack. The setup (Fig. 4) utilizes an Arduino, a DC–DC regulator, voltage/current sensors, and an ADC to emulate the charger. Meanwhile, a Raspberry Pi runs the EVCS server and communicates with the open-source SteVe CMS [37] via WebSockets. Load requests are forwarded from the EVCS to the CMS, where transactions are authenticated and logged. Using Metasploit on Kali Linux, we employed tools such as Ettercap, Bettercap, Nmap, and ARP spoofing to intercept and alter EVCS–CMS traffic, successfully reading and modifying key parameters. Although we did not deploy the full RL-generated attack vectors, this prototype confirms the practical feasibility of FDI attacks in OCPP-based EVCS networks.

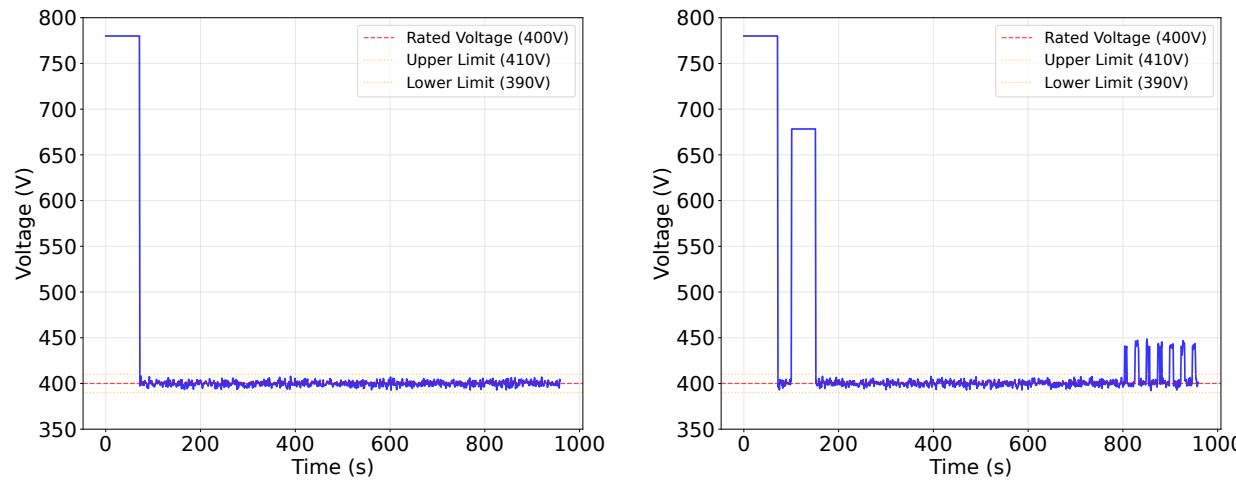

(a) (b)

**Figure 5: Voltage response of sample EVCS output voltage in (a) benign and (b) under attack conditions.**

We have considered OCPP version 1.6 (Fig. 4), which is widely used and does not mandate transport encryption or mutual authentication; therefore, it is often deployed over plaintext WebSocket or server-only TLS. Our attacks, therefore, target the data plane under these common configurations, where messages can be observed and modified in transit. By contrast, deployments using OCPP 2.0.1 security profiles or OCPP 1.6 with correctly configured mutual TLS, provide end-to-end confidentiality and integrity; in such cases, an adversary would first require endpoint compromise or access to trusted keys to decrypt, alter, and re-encrypt traffic before injection of this adaptive adversarial attack is feasible [38, 39]. However, we have kept this part of the scope for this work.

## 4.2 Detailed Analysis Stealthy FDI Attack and Cascading Impact

*4.2.1 EVCS Voltage Fluctuation .* Fig. 5 compares the output voltage of a representative EVCS under normal operation and during an FDI attack. In the benign scenario, voltage trajectories remain within the nominal operational envelope, reflecting stable controller performance. Under FDI conditions, malicious reference injections induce notable voltage deviations, including transient spikes and drift outside safety margins. These fluctuations demonstrate how the attack compromises the CMS's ability to maintain voltage regulation, potentially leading to equipment stress, miscoordination with feeder-level protection, or service interruptions, especially when deployed persistently or across multiple nodes.

*4.2.2 Distribution System Load Fluctuation .* Fig. 6 illustrates the aggregate load profile of the distribution system before and during an FDI attack. The benign profile shows typical daily demand variation shaped by normal EV charging cycles. Under attack, false load, SOC, and voltage signals propagated by compromised CMS instances cause erratic load swings, disrupting feeder balance and potentially triggering abnormal transformer tap operations or thermal violation alarms. These induced fluctuations challenge both operator situational awareness and automated response mechanisms, increasing grid vulnerability and operational costs.

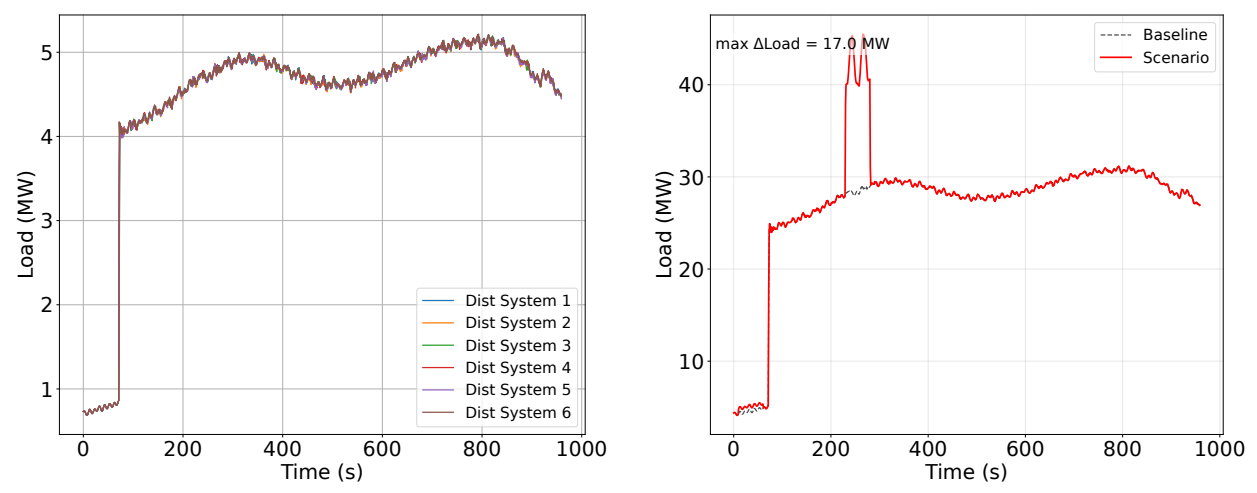

(a) (b)

**Figure 6: Demonstrating distribution grid load demand in (a) benign and (b) disturbance under attack conditions.**

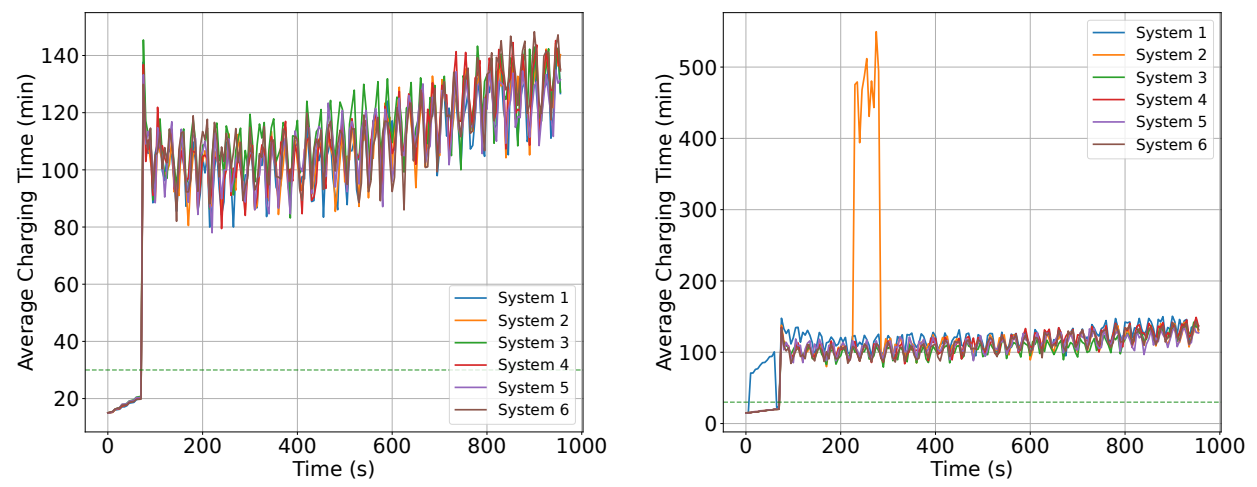

(a) (b)

**Figure 7: Impact on EVCS charging time in (a) benign and (b) under attack conditions.**

*4.2.3 Impact on Average Charging Time.* Fig. 7 presents the distribution of average charging session durations under normal and attack conditions. Nominally, charging times cluster tightly around design targets. When FDI perturbations are injected, session durations scatter and frequently extend beyond acceptable service windows, especially during coordinated attacks. This increase is attributed to undercharging, delayed start, or premature session termination caused by the manipulation of SOC and demand signals, directly degrading user experience and station throughput in affected regions.

*4.2.4 Impact on Average Queue Length.* Fig. 8 shows the temporal evolution of customer queue length at a sample EV charging site. Under normal operation, queue dynamics reflect expected patterns with manageable wait times. Under FDI attack, perturbed charging references and increased charging duration amplify queue lengths and introduce volatility, resulting in longer customer delays, possible service denial, and diminished station utility. These operational bottlenecks can propagate, leading to customer dissatisfaction and reduced grid flexibility to accommodate new charging requests.

*4.2.5 Impact on AGC Reference Power.* Even though we are injecting our FDI attacks into the measurement values used by the CMS at the distribution grid, Fig. 9 demonstrates how FDI attacks at the distribution level can propagate, distorting the load observed by the transmission grid. Normal reference power tracking ensures stable power export/import profiles to the bulk system. Coordinated CMS attacks induce oscillations and setpoint deviations, jeopardizing frequency regulation and reliability margins, particularly during periods of high demand or system contingencies.

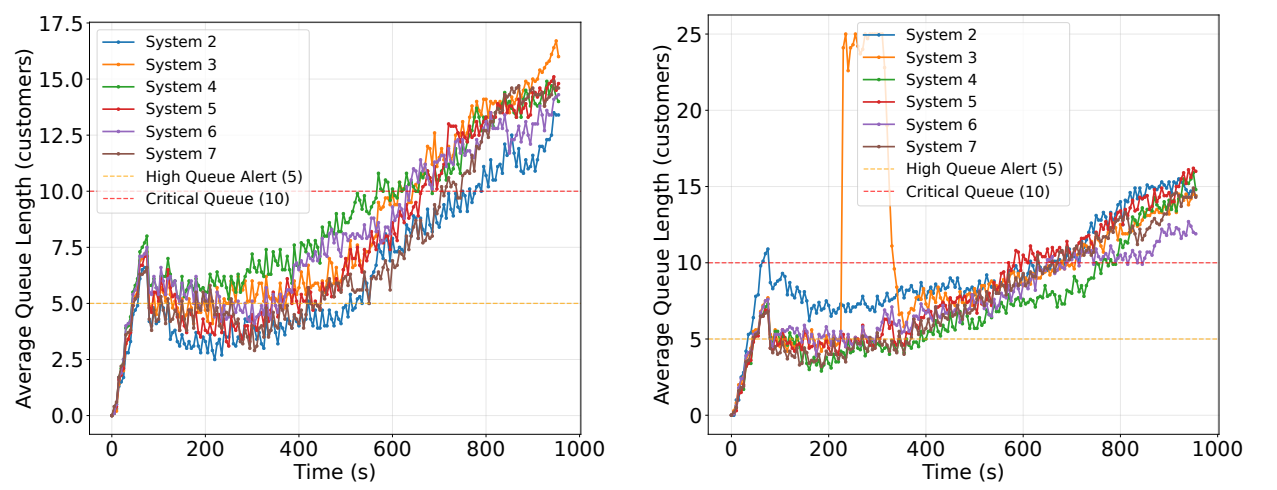

(a) (b)

**Figure 8: Demonstrating average customer queue length in (a) benign and (b) under attack scenarios.**

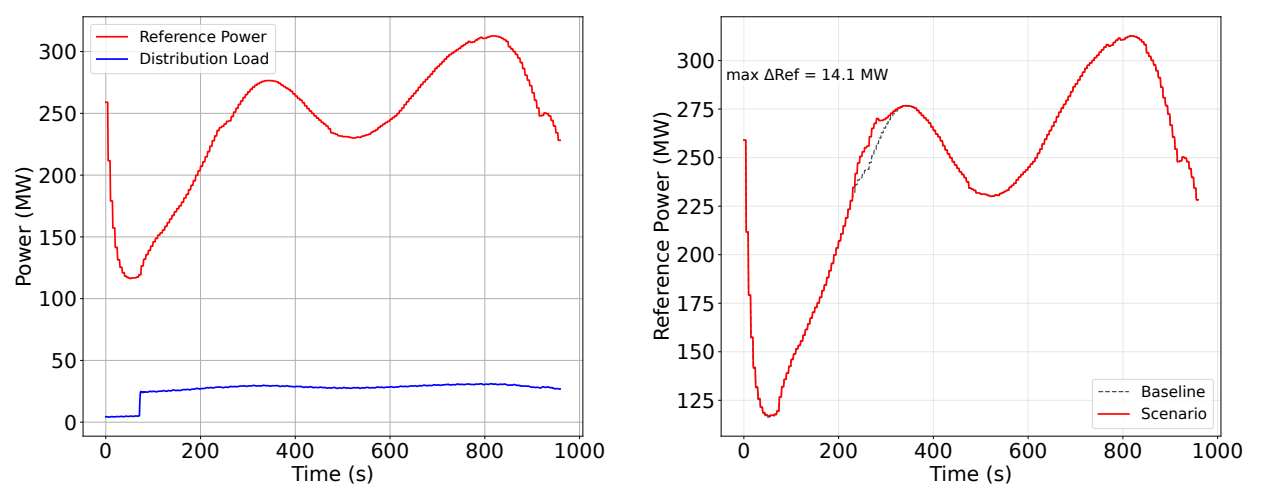

(a) (b)

**Figure 9: Change of reference power of AGC in (a) benign and (b) under attack.**

*4.2.6 Impact on Grid Frequency.* Grid frequency is a key stability indicator, as it is sensitive to unbalanced or rapid changes in load. Under benign conditions, frequency remains tightly regulated around its nominal value (Fig. 10(a)), showing only minor fluctuations. FDI attacks that overwhelm CMS controls or coordinate across multiple stations induce pronounced deviations and oscillations, potentially exceeding governor and underfrequency load shedding thresholds. Fig. 10(b) illustrates small disturbance scenarios that highlight the attack's potential to compromise system-wide frequency stability, posing a risk of cascading events if inadequate mitigation measures are implemented.

*4.2.7 Evaluation of PHANTOM Against LSTM-based ADM.* To systematically evaluate and stress-test this multi-layered defense, the adversarial RL agents, utilizing a hybrid DQN and SAC architecture, learn to craft sophisticated evasion strategies through adversarial training. The DQN agent, operating in a discrete action space, learns to make high-level strategic decisions, including attack type selection (demand increase/decrease, voltage/frequency spoofing, SOC manipulation, and oscillating demand), target system identification, timing window optimization, and evasion strategy selection. Simultaneously, the SAC agent operates in a continuous action space to maximize the attack's impact while minimizing the detection probability. Through this adversarial co-evolution, the agents discover sophisticated evasion tactics that exploit the fundamental blind spots between detection layers: (1) maintaining attack magnitudes just below rule-based thresholds, (2) generating attack patterns that align with statistical distributions of normal traffic to evade anomaly detection, and (3) crafting temporal sequences with carefully controlled autocorrelation structures that confuse the LSTM classifier. Our experiments show that, even in the presence of the LSTM-based ADM, the MARL agents eventually learn fully evasive strategies (as shown in Fig. 11(a)). This stealth comes at a cost: RL training time increases by roughly a factor of five, and the achievable physical impact is reduced. In particular, the maximum frequency deviation under the LSTM-based ADM is limited to about 0.001 Hz as demonstrated in Fig. 11(b), compared to approximately 0.0025 Hz for the rule-based ADM presented in Fig. 10(b), highlighting both the enhanced resilience of the learning-based detector and the more challenging attack exploration landscape it induces.

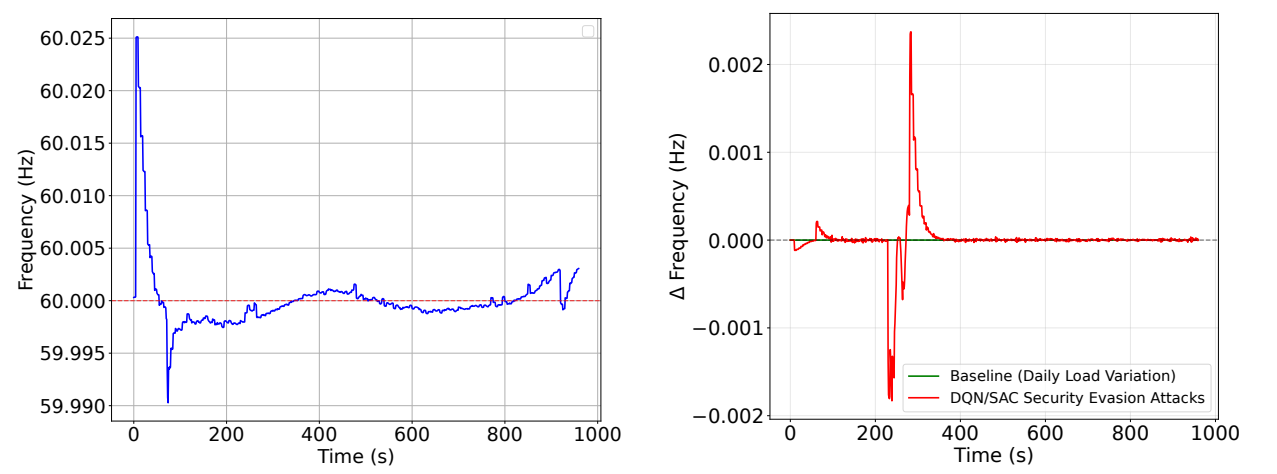

(a) (b)

**Figure 10: Overall frequency response in (a) benign and (b) deviation of frequency due to the attack.**

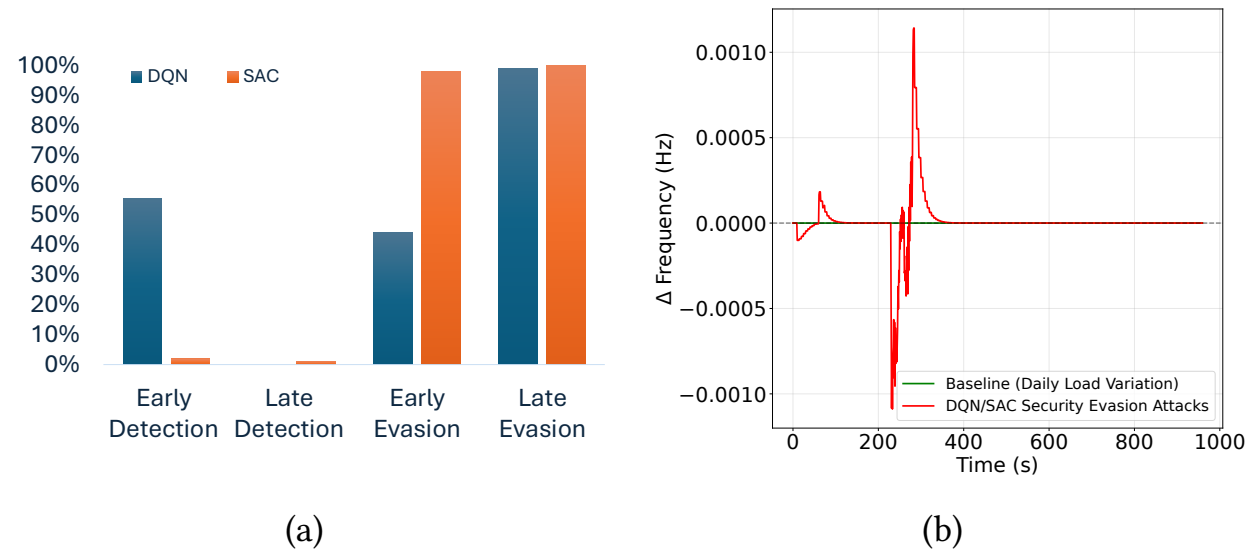

**Figure 11: Demonstrating (a) performance evasion performance of MARL-based attack analytics and (b) frequency deviation of the system in the presence of LSTM-based ADM.**

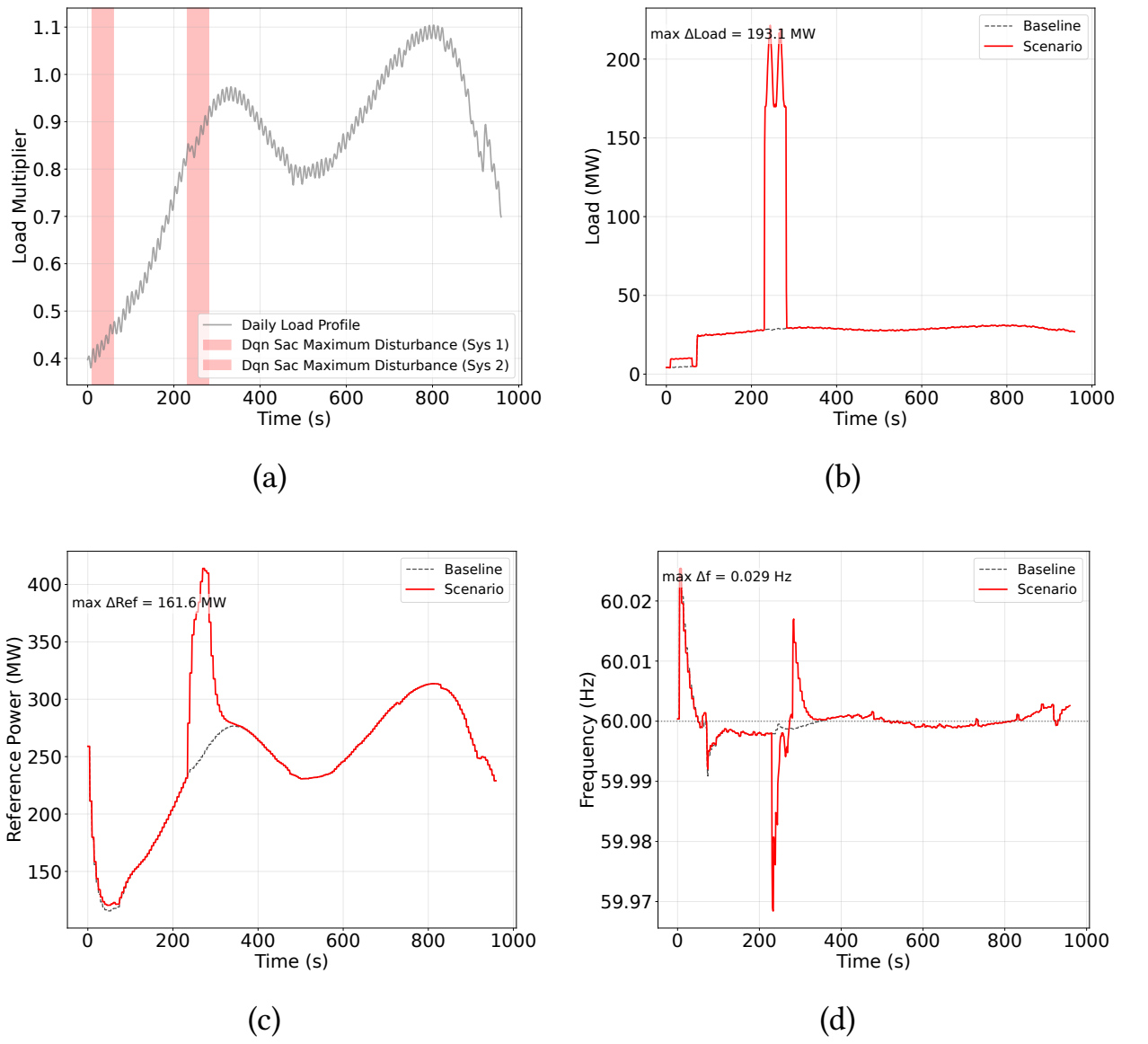

**Figure 12: Demonstrating the impact of FDI attack on CMS without the presence of ADM constraints in comparison with the baseline scenario. (a) Load multiplier with selected area and duration, (b) total distribution load, (c) reference power of AGC, and (d) frequency response of the grid.**

## 4.3 Impact Analysis of FDI Attack Without Intelligent ADM Module

The CMS implements a multi-layered anomaly detection framework that monitors EVCS operational parameters through statistical thresholds and pattern recognition. The ADM continuously tracks charging power deviations, queue length anomalies, and frequency response patterns using baseline comparisons with configurable tolerance bands mentioned in (21)-(28). The ADM employs temporal windowing for trend analysis and adaptive thresholds tuned to historical load profiles and daily variations. The hybrid adversarial agents learn to exploit these boundaries by generating subtle, continuous measurement perturbations that stay below statistical thresholds while maximizing cumulative impact. Reward shaping balances effectiveness and stealth, encouraging attacks that are distributed across time and EVCS stations so that individual anomalies remain weakly correlated and difficult to aggregate into a coherent attack signature. To illustrate the system's vulnerability in the absence of robust anomaly detection mechanisms, we repeat the FDI attack scenarios without any active ADM in place. Disabling the ADM permits the attacker to inject large-magnitude false data directly into the CMS and grid, causing significantly larger deviations in key system variables, such as voltage, load, and frequency, compared to the constrained cases. Without high-fidelity detection and response, attacks persist unchecked—leading to significant destabilization of the EVCS network, prolonged voltage and load recovery, and more severe cascading effects on both distribution and transmission grid operations. This analysis highlights the indispensable role of intelligent ADM schemes for cyber-resilient grid operation under coordinated FDI attacks. The results in Fig. 12(a)-Fig. 12(d) demonstrate significant system-wide impacts of coordinated FDI attacks on the transmission grid. The frequency plot reveals substantial deviations with a maximum excursion of 0.036 Hz below nominal, indicating severe system stress that approaches critical stability thresholds.

## 5 Conclusion

This paper presented PHANTOM-RL, a physics-aware multi-agent reinforcement learning framework for crafting stealth-constrained FDI attacks against federated LSTM–PINN charging management in EVCS-integrated grids. The LSTM–PINN provides physics-consistent real-time setpoints, while a coordinated DQN–SAC attacker learns target selection, timing, and multi-signal perturbations under feasibility and detectability constraints. Hierarchical T&D co-simulation revealed that adaptive FDI can disrupt voltage regulation, increase load, degrade frequency stability, and compromise charging quality, exposing cascading vulnerabilities that extend beyond static or linear analyses. These findings underscore the crucial role of CMS and the necessity for physics-based validation, advanced anomaly detection, and secure federated learning to enhance the resilience of EVCS-connected distribution networks. Future work will prioritize the development of decentralized, self-healing control architectures capable of neutralizing coordinated multi-agent threats in real-time.

## Acknowledgement

This work is supported by the Department of Energy (DOE) under Award# DE-CR0000024. Any opinions, findings, conclusions, or recommendations expressed in this material are those of the authors and do not necessarily reflect the views of the DOE.

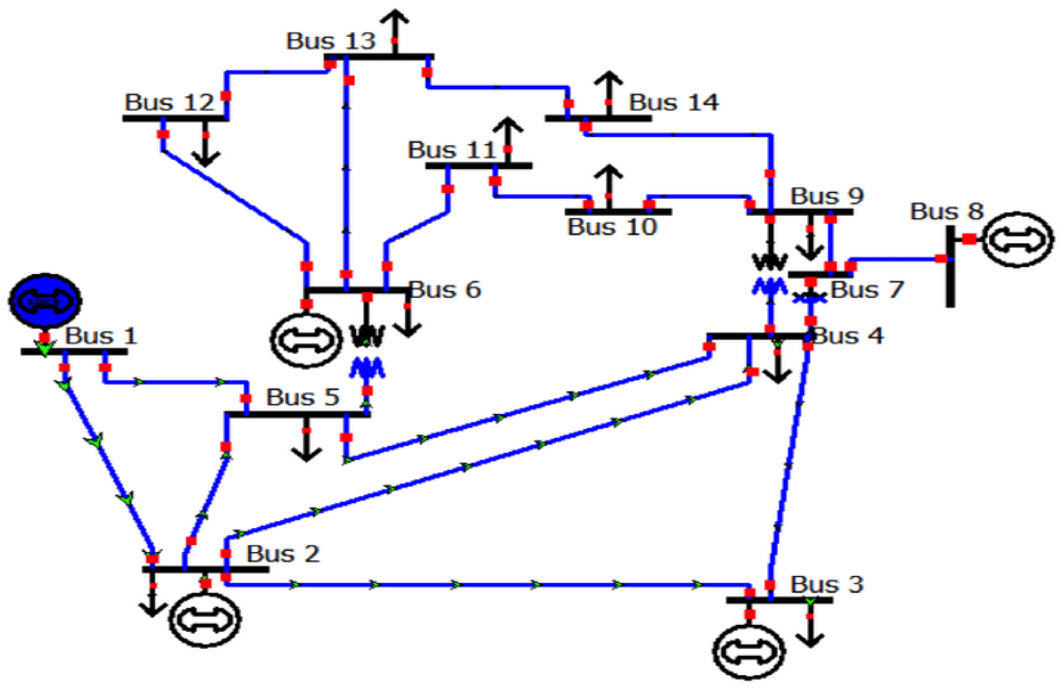

**Figure 13: IEEE 14 Bus for Transmission Grid.**

# A Appendix A

## A.1 Simulation Components

## A.2 IEEE 14 Bus Transmission Grid

The IEEE 14-bus (Fig. 14) test system is a compact transmission benchmark featuring 14 buses, 5 generators, and 11 loads, with roughly 20 branches including several off-nominal transformer taps on a 100 MVA, 60 Hz base. Its topology combines a meshed core (buses 1–7) with a radial tail (buses 9–14), supporting realistic studies of voltage regulation, reactive support, and observability.

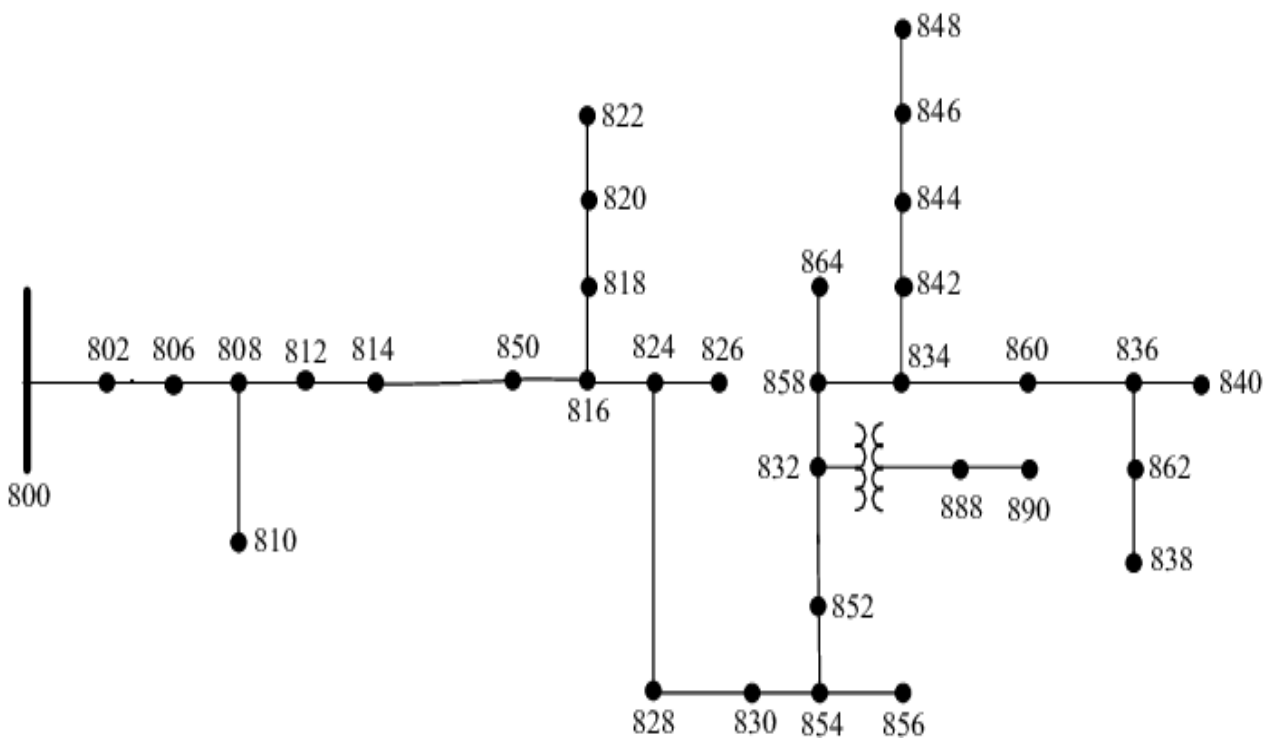

**Figure 14: IEEE 34 Bus for Distribution Feeder.**

## A.3 IEEE 34 Bus Distribution Feeder

The IEEE 34-bus distribution test feeder is a long, radial four-wire wye-grounded medium-voltage circuit widely used to benchmark unbalanced three-phase power flow, volt–VAR control, protection coordination, and DER integration studies. It includes 34 nodes with mixed three-phase and single-phase laterals, spot and distributed ZIP loads, configuration-based line models (impedance/capacitance in ohms/mile), inline voltage regulation via a substation LTC, and on-feeder step-voltage regulators, and switched shunt capacitors. Nominal voltage is 24.9 kV (often scaled to 12.47 kV), with high R/X ratios that accentuate voltage drop, losses, and phase unbalanced conditions.

## A.4 Power Flow Equations and Automatic Generation Control

The power flow equation for active and reactive power can be written for $b^{th}$ bus, which is connected with adjacent $c^{th}$ buses as:

$$\forall_{b,c\in\mathcal{B}^{GV}}\mathcal{P}_g^b = \sum_{c=1}^{c=N}\mathcal{V}_t^{b^2}G_A^{bc} - \mathcal{V}_t^b\mathcal{V}_t^c(G_A^{bc}\cos\delta^{bc} + B_A^{bc}\sin\delta^{bc}) \tag{46}$$

$$\forall_{b,c\in\mathcal{B}^{GV}}Q_g^b = \sum_{c=1}^{c=N}-\mathcal{V}_t^{b^2}(B_A^{bc} + B_{A,0}^{bc}) + \mathcal{V}_t^b\mathcal{V}_t^c(B_A^{bc}\cos\delta^{bc} - G_A^{bc}\sin\delta^{bc}) \tag{47}$$

Automatic generation control of a transmission grid is responsible for keeping the system frequency within a nominal level with the help of tie-line flow measurements, frequency, and generation data obtained from SCADA infrastructure. The AGC uses tie-line flow measurements, frequency, and SCADA generation data to adjust system generation in response to load changes, maintaining the grid frequency at 60 Hz. It regulates frequency and tie-line power flows, ensuring they align with scheduled agreements between neighboring control areas. Based on VG principles, the tie-line power flow can be expressed as:

$$\forall_{ij\in\mathcal{Z}}\Delta\mathcal{P}_{tie}^{ij} = \mathcal{P}_s * (\Delta\delta_{Area_i} - \Delta\delta_{Area_j}) + \mathcal{P}_s * (\Delta\mathcal{V}_{t-Area_i} - \Delta\mathcal{V}_{t-Area_j}) \tag{48}$$

Where,

$$\Delta\delta_{Area}^{ij} = \frac{\sum_i^n \Delta\delta^i}{n} - \frac{\sum_j^m \Delta\delta^j}{m}$$

$$\Delta\mathcal{V}_{t-Area}^{ij} = \frac{\sum_i^n \Delta\mathcal{V}_t^i}{n} - \frac{\sum_j^m \Delta\mathcal{V}_t^j}{m}$$

and $P_s$ is synchronizing power coefficient. When there is a change required in generated power $\mathcal{P}_g^b$, AGC changes the reference set point using a proportional-integral (PI) controller. The dynamics of reference set point and area control error (ACE) determined by AGC for the generators can be written as:

$$\forall_{b,i\in\mathcal{B}^{GV,Z}}\mathcal{P}_{ref}^b = \mathcal{P}_g^b - \int \mathcal{K}^b * \mathcal{ACE}^i \tag{49}$$

$$\forall_{b,ij\in\mathcal{B}^{G,Z}}\mathcal{ACE}^i = \Delta\omega^b * (\frac{1}{R^b} + D^b) + \sum\Delta\mathcal{P}_{tie}^{ij} \tag{50}$$

The governor model we have considered is TGOV1, which is the simplified representation of the steam governor can be expressed by (51):

$$\forall_{b\in\mathcal{B}^{GV}}\mathcal{P}_m^b = \frac{1}{T^b}\int(\frac{\mathcal{P}_{ref}^b - \Delta\omega^b}{R^b} - \mathcal{P}_m^b) \tag{51}$$